\documentclass[12pt]{article}

\newtheorem{dfn}{Definition}

\usepackage{latexsym}
\usepackage{amssymb}
\usepackage{amsfonts}
\usepackage{amsmath}
\begin{document}

\font\mybb=msbm10 at 11pt \font\mybbb=msbm10 at 17pt
\def\bb#1{\hbox{\mybb#1}}
\def\bbb#1{\hbox{\mybbb#1}}
\def\D {\bb{D}}
\def\I {\bb{I}}
\def\R {\bb{R}}
\def\T {\bb{T}}
\def\tr{\rm{tr}}

\newtheorem{proposition}{Proposition}[section]
\newtheorem{theorem}{Theorem}
\newtheorem{lemma}[theorem]{Lemma}
\newtheorem{corollary}[theorem]{Corollary}
\newtheorem{conjecture}[theorem]{Conjecture}
\newtheorem{definition}[theorem]{Definition}
\newtheorem{example}[theorem]{Example}
\newtheorem{condition}{Condition}
\newtheorem{main}{Theorem}
\newtheorem{assumption}[theorem]{Assumption}
\renewcommand{\theequation}{\arabic{section}.\arabic{equation}}
\setlength{\parskip}{\parsep} \setlength{\parindent}{0pt}

\def \outlineby #1#2#3{\vbox{\hrule\hbox{\vrule\kern #1%
\vbox{\kern #2 #3\kern #2}\kern #1\vrule}\hrule}}%
\def \endbox {\outlineby{4pt}{4pt}{}}%
\newenvironment{proof}
{\noindent{\bf Proof\ }}{{\hfill \endbox }\par\vskip2\parsep}
\newenvironment{pfof}[2]{\removelastskip\vspace{6pt}\noindent
 {\it Proof  #1.}~\rm#2}{\par\vspace{6pt}}

\hfuzz12pt

\newcommand{\Section}[1]{\setcounter{equation}{0} \section{#1}}
\newcommand{\cl}[1]{{\mathcal{C}}_{#1}}
\newcommand{\var}{{\rm{Var\;}}}
\newcommand{\cov}{{\rm{Cov\;}}}
\newcommand{\tends}{\rightarrow \infty}
\newcommand{\ep}{{\mathbb {E}}}
\newcommand{\pr}{{\mathbb {P}}}
\newcommand{\re}{{\mathbb {R}}}
\newcommand{\vc}[1]{{\bf {#1}}}
\renewcommand{\vec}[1]{\vc{#1}}
\newcommand{\ra}[2]{{#1}^{(#2)}}
\newcommand{\Z}{{\mathbb {Z}}}
\newcommand{\K}{{\mathbb {K}}}
\newcommand{\ent}{{\rm E}}
\newcommand{\bin}[2]{\binom{#1}{#2}}
\newcommand{\ov}[1]{\overline{#1}}
\newcommand{\diy}{{\displaystyle}}
\newcommand{\uk}{{\underline k}}

\title{\bf Classical feedback for quantum channels}

\vskip 0.5 truecm

\author{Andrew Skeen \\ {\sl Centre for Mathematical Sciences,}\\
{\sl Wilberforce Road, Cambridge, CB3 0WB, UK}\\ email:
ags29@cam.ac.uk}

\maketitle
 \vskip 0.5 truecm
\begin{abstract}
In this paper we investigate whether the use of a noiseless,
classical feedback channel will increase the capacity of a quantum
discrete memoryless channel to transmit classical information.
This problem has been previously analyzed by Bowen and Nagarajan
\cite{bowena} for the case of protocols restricted to product
input states. They showed that feedback did not increase the
information capacity. In this paper we introduce a quantum
analogue of classical causality \cite{massey},\cite{tatikonda} and
prove a capacity theorem (in regularized form) for the
transmission of classical information.

\end{abstract}


\section{Introduction:}

In classical information theory a noiseless feedback channel
between sender and reciver will not increase the Shannon capacity
of a channel. In the quantum case, the situation is more complex
because there are number of possible feedback capacities
corresponding to any channel: the simplest of these is the
product-state input or HSW capacity of a channel aided by
feedback. For this case, Bowen and Nagarajan have shown that there
is no capacity increase over the no-feedback case. On the other
hand, for the case of quantum capacities, it has been shown by
Bowen \cite{bowenb} that the use of a classical feedback channel
may increase the value of the channel capacity from $Q$ to $Q_E$ ,
where $Q$ is the quantum capacity of the channel, and $Q_E$ is
entanglement-assisted capacity of the channel (see \cite{bennett2}
for rigorous definitions). In this work, we will concentrate on
classical feedback and the effect on the "full" classical capacity
of the channel (in which case entangled input states are allowed).
In this regard it is worth referring to an important conjecture of
quantum information theory,
 that the unassisted capacity, called $C$,
is in fact additive: $$C(\Phi\otimes\Psi)=C(\Phi)+C(\Psi)$$ for
any two quantum channels $\Phi$ and $\Psi$ \cite{holevo}. It is
natural to conjecture whether a similar additivity property holds
in the case of (suitably defined) classical feedback capacity
(which in the sequel, will be denoted $C_F$). If such a conjecture
were true, then by the result of Bowen \cite{bowena} it would
follow that unconstrained classical feedback would not increase
the channel capacity. However, this turns out not to be the case,
as Devetak and co-workers \cite{dev} have produced an example of a
discrete memoryless quantum channel for which the feedback
capacity exceeds the Holevo product state capacity. In this paper
we will give a coding theorem for quantum discrete memoryless
channels with classical feedback and demonstrate that, as in the
classical case, the definitions and proofs are highly dependent on
an analog of the classical notion of causality introduced in
\cite{massey}.

\vskip 0.5 truecm

In what follows we make use of standard notation for information
transfer through quantum channels: a quantum channel $\Phi$ with
input space ${\cal H}$ is modelled as a trace-preserving
completely positive map on density matrices  $\rho\in{\cal
B}({\cal H})$ and will be represented by a Kraus decomposition
$\{E_{j}\}$ with $\sum_j E_j^*E_j=I$ as $\Phi(\rho)=\sum_j E_j\rho
E_j^*$. In the direct part of the coding theorem proof below we
will consider encoding using density operators picked from
ensembles of the form $\{p_x,\rho_x\}$, where $p_x$ is the
probability of picking density matrix $\rho_x$.

We will work with a quantum analogue of classical mutual
information, which we define for a bipartite quantum system
$\rho_{{\cal A}{\cal B}}$, by $$I({\cal A}:{\cal B})=S(\rho_{{\cal
A}})+S(\rho_{{\cal B} })-S(\rho_{{\cal A}{\cal B}}),$$ where
$S(\sigma)$ is the von Neumann entropy of the density matrix
$\sigma$. This quantity was introduced by Adami and Cerf
\cite{adami}. \vskip 0.5 truecm In particular we will consider
states of the form $$\rho_{A{\cal B}}=\sum_{i}p_i |i\rangle\langle
i|\otimes \rho_i^B,$$ where $\{|i\rangle\}$ is an orthonormal
basis for subspace $A$ (for clarity we will generally represent
quantum subsystems with calligraphic letters as opposed to
ordinary capitals for classical registers). Such a state is said
to exhibit ``classical-quantum'' correlations and we have the
following form for $I(A:{\cal B})$: $$I(A:{\cal B})=S\big(\sum_i
p_i\rho_i\big)-\sum_i p_iS(\rho_i)=\chi({\cal E}),$$ where ${\cal
E}=\{p_i,\rho_i\}$ and $\chi({\cal E})$ is the Holevo quantity of
ensemble ${\cal E}$.

The quantum mutual information allows us to more neatly express
the quantities of interest in classical information transfer.

\section{ Feedback Code Definition and feedback capacity upper bound}

What follows is a formulation of the feedback communication
protocols similar to that introduced by Bowen and Nagarajan in
their paper \cite{bowena}. As before, the message source is a
finite alphabet stochastic process satisfying the asymptotic
equipartition property (AEP) , see \cite{shan}.

Given a rate $R>0$, we define an $n$-block feedback code of size
$N=2^{nR}$ for channel $\Phi$ acting on states in the input
Hilbert space ${\cal H}$ as a quadruple ${\cal C}_n=({\cal
E}_{F},{\cal M}_{F},{\cal N}_{F},f)$ consisting of \vskip 0.5
truecm 1) strings $i_1^n(l)\in{\cal A}^n$ forming a classical code
${\cal C}_{N}$ of size $N$. These strings should be viewed as the
elements of the image of a mapping from the message space ${\cal
M}$ to the space of input strings. \vskip 0.5 truecm 2) an input
ensemble ${\cal E}=\{p_{i_1^n(l)},\rho_{i_1^n(l)}\}$ where each
$\rho_{i_1^n(l)}$ is a density matrix in ${\cal H}^{\otimes n}$.
For convenience we will denote this Hilbert space as
$\otimes_{k=1}^n{\cal H}_k$, with the index $k$ referring to the
space on which the $k$-th sequential channel action occurs. \vskip
0.5 truecm 3) a collection of measurements ${\cal M}_{F}$ given by
measurements ${\cal M}_1,\cdots,{\cal M}_n$, where ${\cal M}_j$
acts in the space ${\cal L}({\cal H}_1)\otimes\cdots\otimes{\cal
L}( {\cal H}_j),\,1\leq j\leq n$, and an array ${\cal N}_{F}$ of
"associated" trace preserving completely positive maps $\{{\cal
N}_{2}^{(k_1)}\},\cdots,\{{\cal N}_{n}^{(k_{n-1})}\}$, where $k_i$
runs through the outcomes of measurement $M_i$. Also ${\cal
N}_{j}^{(k_{j-1})}$ acts on the space
$\displaystyle\otimes_{k=j+1}^{n}{\cal H}_{k},\,1\leq j\leq n-1$.
We denote the elements of ${\cal N}_{j}^{(k_{j-1})}$ by
$\{N_{j,l}^{(k_{j-1})}\}$ and those of
 ${\cal M}_i$ by $\{F_{k}^i\}$. For clarity, assume that the outcomes
$k_n$ of ${\cal M}_n$ are strings $i_1^n(l)\in{\cal C}_n$ and a
splodge (error) denoted "er". \vskip 0.5 truecm


The transmission protocol is then to sequentially transmit the
codeword by uses of the channel , at each stage measuring the
state received so far and using a noiseless classical channel to
transmit the measurement outcome to the sender. Formally: the
first round of communication starts with the mapping:
$\rho_{i_1^n(l)}\mapsto\omega_{i_1^n(l)}^0:=(\Phi\otimes
I\otimes\cdots\otimes I)(\rho_{i_1^n(l)})$. The feedback
measurement ${\cal M}_1$ is made and the outcome $k_1$ transmitted
to the sender who then applies map  ${\cal N}_{2}^{(k_1)}$ to the
state
$$\frac{F_{k_1}^{1}\omega_{i_1^n(l)}F_{k_1}^{1*}}{{\rm{tr}}(F_{k_1}^{1}\omega_{i_1^n(l)}F_{k_1}^{1*})}$$
Here we use the notation $F_{k_1}$ for $F_{k_1}\otimes I\otimes
\cdots\otimes I$ and a similar agreement holds in what follows.
The result of these operations is the state:
$$\omega^{1}(i_1^n(l),k_1)=\sum_{j}\frac{\big(N_{2,j}^{(k_1)}(F_{k_1}^{1}\omega_{i_1^n(l)}F_{k_1}^{1*})N_{2,j}^{(k_1)*}\big)}{{\rm{tr}}(F_{k_1}^{1}\omega_{i_1^n(l)}F_{k_1}^{1*})}$$

We proceed inductively, with
$\omega^{m-1}(i_1^n(l),k_1^{m-1})$ obtained from
$\omega^{m-2}(i_1^n(l),k_1^{m-2})$ by
$$\omega^{m-1}(i_1^n(l),k_1^{m-1})=\sum_{j}\frac{\big(
N_{m,j}^{(k_{m-1})}(F_{k_{m-1}}^{m-1}(I\otimes\cdots\Phi\otimes\cdots
I)\omega^{m-2}_{i_1^n(l)}((k_1^{m-2}))F_{k_{m-1}}^{m-1
*})N_{m,j}^{(k_{m-1})*}\big)}
{{\rm{tr}}(F_k^{m-1}(I\otimes\cdots\otimes\Phi\otimes\cdots
I)\omega^{m-1}_{i_1^n(l),}((k_1^{m-2}))F_k^{(m-1)*})}.$$

For the fidelity of this procedure we consider random outcomes
$K_j$ for every measurement ${\cal M}_j,\,j=1,\cdots,n$. The final
(random) estimate of the original classical string is then a fixed
function $f(K_1,\cdots,K_n)$ taking values in ${\cal C}_N$.

Define the error probability for this code as: $${\bf P}_{{\cal
E},{\cal M}_{\rm{F}},{\cal N}_{\rm{F}}}
=1-\max_{l}[p(i_1^n(l)){\mathbf
P}(f(K_1,\cdots,K_n)=i_1^n(l)|i_1^n(l))]$$ then take the minimum
over all codes $({\cal E}, {\cal M}_{\rm{F}},{\cal N}_{\rm{F}})$:
$${\bf P}_{e}(n,N)=\min\;{\bf P}_{{\cal E},{\cal M}_{\rm{F}},
{\cal N}_{\rm{F}}}.$$ The rate $R$ is achievable if
$\lim_{n\to\infty}{\bf P}_{e}(n,N)=0$. The feedback capacity
$C_{\rm{F}}$ is then defined as the supremum of all achievable
rates.

\vskip 0.5 truecm These operations can be summarised in the
quantum mutual information formalism by defining a sequence of
extended Hilbert space quantum states. We begin by defining
$$\rho_{A_1^nX_1^{n-1},{\cal
Z}_1^n}^0=\sum_{l}p_{i_1^n(l)}|i_1(l)\rangle\langle i_n(l)|\otimes
| i_1(l)\rangle\langle i_n(l)|\otimes |0\rangle\langle
0|\ldots\otimes |0\rangle\langle 0|\otimes\rho_{i_1^n(l)}.$$ The
non italicised systems $X_1^{n-1}$ (each in an initial state
$|0\rangle\langle 0|$) and $A_1^n$ are classical registers
recording, respectively, the classical codewords and the outcomes
of the feedback measurements. To achieve this, the POVM elements
of a given feedback measurement are augmented to the form
$U_{k_i}\otimes M_{k_i}$, where $U_{k_i}$ is a unitary operator
acting on the register system $X_i$. By applying the sequence of
operations outlined above we obtain the states:

\begin{align}
\rho_{A_1^nX_1^{n-1},{\cal
Z}_1^n}^t&=\sum_{l}p_{i_1^n(l)}p(k_1|i_1^n(l))\ldots
p(k_{i}|i_1^n(l),k_1\ldots,k_{i-1})\otimes|i_1(l)\rangle\langle
i_n(l)||i_n(l)\rangle\langle i_n(l)| \notag\\ & \quad\otimes
| k_1\rangle\langle k_1|\ldots\otimes |k_{i-1}\rangle\langle
k_{i-1}|\ldots|0\rangle \langle
0|\otimes\omega(i_1^n(l),k_1^{i-1})^t,
\end{align}

 for $1\leq
t\leq n-1$.

Now after $k$ rounds of communication, the state held by the
receiver can be written in the EHS form as ${\rm
{tr}}_{A_{k+1}^nX_k^{n-1}{\cal
Z}_{k+1}^n}\big(\rho_{A_1^nX_1^{n-1},{\cal Z}_1^n}^k \big)$ This
reflects the fact that the ensemble held by the receiver at this
point contains states (of the form ${\rm
{tr}}_{A_{1}^nX_k^{n-1}{\cal Z}_{k+1}^n}
\big(\rho_{A_1^nX_1^{n-1},{\cal Z}_1^n}^k \big)$) that can be
labelled by strings of length $k$ from the alphabet ${\cal A}$, or
equivalently, indexed by the "register" space $A_1^k$. Taking the
partial trace with respect to $A_{k+1}^n$ leaves us with an EHS
state indexed by this register. This can be viewed in an analogous
way to the classical causal systems (introduced by
Massey \cite{massey}): A feedback protocol such as that defined in
this chapter implies the existence of a 
classical-quantum Markov chain \cite{windev}
$M_1^n\to A_1^n\to {\cal Z}_1^n$. This in turn implies that
\small{\begin{equation}\begin{array}{ll}
P(K_n=k_n|K_1^{n-1}=k_1^{n-1},X_1^n=x_1^n,M_1^k=m_1^k)&=P(K_n=k_n
| K_1^{n-1}=k_1^{n-1},X_1^n=x_1^n)
\end{array}\end{equation}}
The operational 
interpetation, as in the classical case, is that the message is
specified before the initial transmission encoding and the channel
is only aware of the message identity via its past inputs,
measurement outputs and current input.
\begin{dfn}
For a sequence of EHS states corresponding to a $n$-block feedback code,
as defined above, we have the {\it quantum directed information}, given
by the formula
$$I(A_1^n\rightarrow {\cal Z}_1^n)=\sum_{t=1}^n I_t(A_1^t:{\cal Z}_t|
{\cal Z}_1^{t-1}),$$
where the notation $I_t$ refers to the mutual information
calculated with respect to the EHS state $\rho_{A_1^nX_1^{n-1},{\cal
Z}_1^n}^t\;,\leq t\leq n$.
Furthermore, we will make use of a related quantity
$$I_n(A_1^n\rightarrow {\cal Z}_1^n)=\sum_{t=1}^n I_n(A_1^t:{\cal Z}_t|
{\cal Z}_1^{t-1}).$$
\end{dfn}

We have the following lemma:

\begin{lemma}{\bf Directed Data Processing Inequality}
With the above definitions, we have the following inequality
$$I(M_1^n:{\cal Z}_1^n)\leq I(A_1^n\rightarrow {\cal Z}_1^n).$$
\end{lemma}
\begin{proof}
We first show that
$$I(M_1^n:{\cal Z}_1^n)\leq I_n(A_1^n\rightarrow {\cal Z}_1^n).$$
 To do this, we
imitate the methods of Massey \cite{massey}:

\begin{equation}\begin{array}{ll}
S({\cal Z}_1^n|M_1^n)&=\sum_{k=1}^{n}S({\cal Z}_k|{\cal Z}_1^{k-1}M_1^n)\\
\,&\geq \sum_{k=1}^{n}S({\cal Z}_k|{\cal Z}_1^{k-1}A_1^k M_1^n)\\
\,&=\sum_{k=1}^{n}S({\cal Z}_k|{\cal Z}_1^{k-1}A_1^k),
\end{array}\end{equation}

where we have used the fact that conditioning reduces the
conditional von Neumann entropy (a direct consequence of strong
subadditivity) and that $M_1^n\to A_1^n\to {\cal Z}_1^n$ is a
classical-quantum Markov chain.
It then follows that
\begin{equation}\begin{array}{ll}
I(M_1^n:{\cal Z}_1^n)&\leq \sum_{k=1}^n I_n(A_1^k:{\cal Z}_k|
{\cal Z}_1^{k-1})\\
&=I_n(A_1^n\rightarrow {\cal Z}_1^n).
\end{array}\end{equation}

Furthermore, we have $I_n(A_1^k:{\cal Z}_k|
{\cal Z}_1^{k-1})\leq I_k(A_1^k:{\cal Z}_k|
{\cal Z}_1^{k-1})\;,1\leq k\leq n$, a consequence of the conditional
version of the data-processing inequality (see appendix), from which the 
result follows.
\end{proof}

\begin{theorem}
Under the conditions described above, we have
$$C_F=\lim\sup_{n\to\infty}\frac{1}{n}I(A_1^n\to{\cal Z}_1^n).$$
Here the supremum is over all classical-quantum states
corresponding to $n$-block feedback codes, subject to the
additional constraint that the above limit exists.
\end{theorem}

\vskip 0.5 truecm The theorem will be proved in 2 parts: Below we
will the demonstrate the converse
$$C\leq\lim\sup_{n\to\infty}\frac{1}{n}I(A_1^n\to{\cal Z}_1^n),$$
and provide a code that asymptotically (with respect to
$n\to\infty$) achieves this upper bound, hencing showing that
$$C\geq\lim\sup_{n\to\infty}\frac{1}{n}I(A_1^n\to {\cal Z}_1^n).$$

\begin{proof}
 Fix $n$ and consider an $n$-block product state feedback code
$${\cal C}_n=\{{\cal E},{\cal M},{\cal N},f\}.$$ Define the
message as $M_1^n$ and $A_1^n$ as the input codeword random
variable with probability distribution ${\bf
P}(A_1^n=i_1^n(l))=p_{i_1^n(l)}$. Our proof will involve
calculating the classical mutual information between this random
variable and the random vector $K_1^n=(K_1,\ldots,K_n)$. The
probability distribution of this random vector is determined by
the following set of relations:
$${\bf P}(K_1=k_1)=\sum_l p_{i_1^n(l)}{\rm
{tr}}(\rho_{i_1^n(l)}F_{k_1}^1 F_{k_1}^{1*})$$ and 

\small{\begin{equation}\begin{array}{ll}
{\bf
P}(K_m=k_m|K_1^{m-1}=k_1^{m-1})&={\rm
{tr}}((I\otimes\cdots\Phi\otimes\cdots\otimes
I)\omega(i_1^n(l),k_1,\ldots,k_{m-1})
F_{k_m}^{m}F_{k_m}^{m*}),
\end{array}\end{equation}}
for $1\leq m\leq n$.

It is sufficient \cite{gallager} 
to provide an upper bound on $H(M)$, the single-letter entropy of
the message source, when ${\bf P}_e(n,2^{nR})\to 0$. This is done
by using the Fano inequality and the classical data-processing
inequality \cite{cover}:

\begin{equation}\begin{array}{ll}
\frac{H(M)}{n}&\leq \frac{H(M|f(K_1^n))+I(M:f(K_1^n))}{n}\\
\,&\leq \frac{1+{\bf P}_e(n,2^{nR})(nR)+I(M:K_1^n)}{n}\\
\,&=\epsilon_n+\frac{I(M:K_1^n)}{n}
\end{array}\end{equation}
where $\epsilon_n\to 0$ as $n\to \infty$.

Now from the Holevo bound \cite{holevo}, or alternatively, the
data processing inequality \cite{ahlswede}, we have
$$I(M:K_1^n)\leq I(M:{\cal Z}_1^n).$$ 
Then applying the directed data-processing inequality, Lemma 2,
we have 
$$I(M:{\cal Z}_1^n)\leq I(A_1^n\to {\cal Z}_1^n).$$

Substituting and letting $n\to\infty$ then gives

$$C_F\leq\lim\sup_{n\to\infty} \frac{I(A_1^n\rightarrow {\cal
Z}_1^n)}{n}$$
\end{proof}
\section{Achievability proof}

In this section we complete the proof of Theorem 3 for discrete
memoryless channels with feedback, by showing the direct part. We
have already proved an upper bound for the feedback capacity using
the Fano inequality. We will show that this bound is achievable by
providing a code which asymptotically achieves this bound (as the
number of channel uses $\to\infty$). The proof proceeds along
similar lines to the HSW theorem proof, using a generalized
version of the square-root measurements used in the direct part of
the proof of that theorem. \vskip 0.5 truecm We will consider the
case of discrete memoryless channels. Our coding procedure is to
use a "double-blocked" code- we will construct an $nl$-block
feedback code from  the simultaneous use of $l$ independent
instances of an $n$-block feedback code. Our approach is
essentially to perform an entangling measurement on the $l$-fold
tensor product of the $k$-th round outputs of each $n$-block
instance, with the purpose of correctly obtaining the classical
strings labelling these states, with high probability. At the same
time, these measurements will be shown to (in a sense defined
later) not disturb the states they act on very much. Finally, we
allow the block parameters $n,l\to \infty$ and obtain the
asymptotic rate achievable using such an encoding and show that
this rate is the upper bound obtained in the converse proof. The
technical details of the approach are described below: \vskip 0.5
truecm

Given a rate $R>0$ and the block-length $nl$, we define our code
of size $N=2^{nlR}$ in terms of $l$ copies of $n$-block feedback
code ${\cal C}_n$ of size $2^{nR}$ : the code is random, and each
codeword  of length $nl$ is built up as a concatenation of the $l$
words $i_1^{n}(j)\;1\leq j\leq l$ chosen at random from the code
${\cal C}_n$. The corresponding quantum codeword is the $l$-fold
tensor product of the codeword states chosen. The communicating
parties then use the following order of transmission and feedback
operations: transmission rounds $kl+1$ to $(k+1)l$ consist of
following the $(k+1)$-th round of each copy of the protocol ${\cal
C}_n$, independently of the other copies
 for $1\leq k\leq n-1$. In addition at round
$j\,l,\;1\leq j \leq n$, the receiver performs a measurement
${\cal R}_j$ with outcome denoted $R_j$, on the state in his
possession at that point. The round $jl$ will, for convenience, be
referred to as the $j$-th ``global round''. The measurement will
identify which state the receiver has transmitted up to that
point. The specific form of these measurements ${\cal R}_j$ will
be discussed in the next section. The result of this measurement
is returned to the sender and if it does not agree with the
classical data sent, the protocol terminates with an error. The
nature of the protocol implies that an error can occur on global
round $j\;1\leq j\leq n$.

If no error occurs, the final outcome of the feedback protocol, is
a function $f(R_1,\ldots,R_l)$ of $(R_1,\ldots,R_l)$, taking
values in the set of input message strings of length $n\;l$.
\vskip 0.5truecm For any choice of message $m$ will denote the
input classical words of this protocol by $i_1^n(1,\ldots,l)$ or
in shortened form $i_1^n({\underline l})$, where this refers to
the concatenation $(i_1^n(1),\ldots,i_1^n(l))$ of the classical
codewords of the $l$ individual original copies of ${\cal C}_n$.
We suppress the dependence on $m$ here, because in the sequel, the
message source will be assumed to have a uniform distribution over
all possible messages, since, by the classical theory, this will
yield the maximal average error probability \cite{gallager}.
\vskip 0.5truecm The initial ensemble held by the sender is then
of the form $$\{p_1^n(1)\ldots
p_1^n(l),\rho_{i_1^n(1)}\otimes\ldots\otimes\rho_{i_1^n(l)}\},$$
reflecting the fact we consider, a priori, $l$
tensor-product/independent input protocols. Equivalently, the
input states for such a protocol can also be expressed in the EHS
form discussed earlier as $\rho_{A_1^nX_1^{n-1}{\cal
Z}_1^{n}}(1)\otimes\ldots\otimes\rho_{A_1^nX_1^{n-1}{\cal
Z}_1^n}(l)=\rho_{A_1^n X_1^{n-1}{\cal Z}_1^n}({\underline l}),$
say, where each term in the tensor product is the EHS
representation of the input protocol for one of the $l$ copies of
${\cal C}_n$. Now the state transmitted in transmission rounds
$(k-1)l+1$ to $k l$ can be labelled with length $l$ strings
$i_k(1,\ldots,l)=i_k({\underline l})$ (in the notation introduced
above). In the EHS representation the state held by the receiver
at this point is given by ${\rm {tr}}_{(A_1^kX_1{\cal
Z}_1^k)}\left(\rho_{A_1^n X_1^{n-1}{\cal Z}_1^n}({\underline
l})\right),$ tracing out all except the first $k$ rounds in each
${\cal C}_n$ sub-protocol. Thus the state held by the receiver at
this time is labelled by the "classical" register $A_1^k$, or
alternatively, by strings $i_1^k({\underline l})$ of length $l\;
k$. Similarly, the state received in rounds $(k-1)l$ to $(k+1)l$
can be labelled by length $l$ strings $i_{k}({\underline l})$. In
our notation, the same strings will label the measurement ${\cal
R}_k$.

\vskip 0.5 truecm

We will give a specification of the code, described generally in
the first section, which achieves the capacity upper bound.

The decoding procedure is defined as follows: For $\delta>0$, let
$\epsilon_k=2^{-lkc\delta^2}$, for some $c>0$ and $1\leq k\leq n$.
We begin by constructing ${\cal R}_1$ as a set of (Hermitian)
operators $R_{r_1}$ that satisfy sub-POVM condition : $$\sum_{r_1}
R_{r_1}\leq I,$$ where $r_1=j_1^n$ is the measurement outcome. For
full generality, we will define $R_0=I-\sum_{j_1^n} R_{j_1^n}$ as
the POVM element pertaining to a faulty output. The form of the
measurements, described below, will depend explicitly on
$k_1({\underline l})$, so the measurement made is in fact
conditional on the feedback measurement outcomes for the
individual copies of ${\cal C}_{n}$.
 The
measurements ${\cal R}_j,\;2\leq\;j\;\leq\;n$ are constructed in a
similar fashion and are conditional on the outcomes of ${\cal
R}_1,\ldots, {\cal R}_{j-1}$ and the feedback measurements of the
individual "copies" of ${\cal C}_n$, which will be represented by
concatenated strings, in a slight abuse of notation, as
$i_1^{j-1}({\underline l})=(i_1({\underline
l}),\ldots,i_{j-1}({\underline l}))$.

Given an initial ensemble as defined above, $${\cal
E}=\displaystyle\{p_{i_1^n(1)}\ldots
p_{i_1^n(l)},\rho_{i_1^n(1)}\otimes\ldots\otimes
\rho_{i_1^n(l)}\displaystyle\},$$

after $tl$ rounds of communication, conditional on no error in
measurements ${\cal R}_1,\ldots, {\cal R}_{t-1}$ and given
sequence of feedback outcomes
$$(k_1^t(1),\ldots,k_1^t(l))=k_1^t({\underline l})$$ (in our
notation), the state held by the receiver is in the ensemble
denoted $${\cal E}(k_1^t({\underline l}),i_1^{t-1}({\underline
l}))= \big\{p\left(i_1^n({\underline l})|i_1^{t-1}({\underline
l}),k_1^t({\underline l})\right),{\rm {tr}}_{(t+1\to
n)}\big(\omega_{i_1^n(l)}^{t}(k_1^t({\underline
l}),i_1^{t-1}({\underline l})\big) \displaystyle\} ,$$ where the
notation ${\rm {tr}}_{(t+1\to n)}$ implies that we are tracing out
all but the first $t$ rounds of communication for each copy of
${\cal C}_n$ considered. To avoid confusion, note that the
$i_1^{t-1}({\underline l})$ upon which we condition, refers, as
noted above, to the string that is the concatenation of the
outcomes of the measurements ${\cal R}_1,\ldots,{\cal R}_t$. This
reflects the fact that, with probability $p\left(i_1^n({\underline
l})|i_1^{t-1}({\underline l}),k_1^t({\underline l})\right)$, the
joint state held by the sender and receiver is denoted
$\omega_{i_1^n({\underline l})}^{t}(k_1^t({\underline
l}),i_1^{t-1} ({\underline l}))$.

Here we define
$\omega_{i_1^n(\underline{l})}^{t}(k_1^t(\underline{l}),i_1^{t-1}(\underline{l}))$
in terms of
$\omega_{i_1^n(\underline{l})}^{t-1}(k_1^{t-1}({\underline
l}),i_1^{t-2}(\underline{l}))$ via the relation
$$\omega_{i_1^n(l)}^{t}(k_1^t({\underline l}),i_1^{t-1}(l))=\otimes_{j=1}^l {\cal N}^{k_{t}(j)}\circ\otimes_{j=1}^l
{\cal M}_{t,k_t(j)}^j\circ
\left(\otimes_{j=1}^l\Phi^{(t)}(j)\right)$$
$${\cal R}_{t-1,i_{t-1}({\underline l})}(i_1^{t-2}({\underline
l}),k_1^{t-1}({\underline l}))\left(\omega^{t-1}_{i_1^n({\underline
l})}(k_1^{t-1}({\underline l}),i_1^{t-2}({\underline
l}))\right),$$ where ${\cal R}_{t-1,i_{t-1}({\underline
l})}(i_1^{t-2} ({\underline l}), k_1^{t-1}({\underline l}))$ is
the realisation of the single element POVM given by 
\small{\begin{equation}{\cal
R}_{t-1,i_{t-1}({\underline l})}(k_1^{t-1}(l),i_1^{t-2}(l))(\rho)
=\frac{\sqrt{R_{i_{t-1}({\underline l})}(k_1^{t-1}({\underline
l}),i_1^{t-2} ({\underline l}))}
\rho\sqrt{R_{t-1,i_{t-1}({\underline l})}(k_1^{t-1}({\underline
l}),i_1^{t-2} ({\underline l}))}} {{\rm{tr}}(\rho
R_{i_{t-1}({\underline l})}(k_1^{t-1}({\underline
l}),i_1^{t-2} ({\underline l})))}
\end{equation}}

for POVM elements $R_{t-1,i_{t-1}({\underline
l})}(k_1^{t-1}({\underline l}),i_1^{t-2} ({\underline l}))$ as
specified below, corresponding to measurement outcome
$i_t({\underline l})$. ${\cal M}_{t,k_t(j)}$ is a realisation of
the single-element POVM corresponding to the feedback measurement
outcome $k_t(j)$ (on $j$-th copy of ${\cal C}_n$), defined, for
input density matrix $\rho$ as:

$$ {\cal M}_{t,k_t(j)}(\rho)=\frac{F^t_{k_t(j)}\rho
F^{t*}_{k_t(j)}}{{\rm {tr}}(\rho F^t_{k_t(j)}F^{t*}_{k_t(j)})} ,$$

Also $\Phi^{(t)}(1)\otimes\ldots\otimes\Phi^{(t)}(l)$ represents
the channel action on the $tl+1\to (t+1)l$ rounds and ${\cal
N}^{k_t(1)} \otimes\ldots\otimes{\cal N}^{k_t(l)}$, the
feedback post-processing done by the sender before transmission on
those rounds. Also, we have inductively

\small{\begin{equation}\begin{array}{l}
p\left(k_t({\underline l} )|i_1^n({\underline
l}),k_1^{t-1}({\underline l}),i_1^{t-1}({\underline l})\right)={\rm{tr}}
\left(\otimes_{j=1}^l\Phi^{(t)}(j)\right)\\
\circ\left({\cal R}_{t-1,i_{t-1}({\underline l})}(i_1^{t-2}({\underline
l}),k_1^{t-1}({\underline l}))\left(\omega^{t-1}_{i_1^n({\underline
l})}(k_1^{t-1}({\underline l}),i_1^{t-2}({\underline
l}))\right)\right)\otimes_{j=1}^l F_{k_t(j)}^jF_{k_t(j)}^{j*}
\end{array}
\end{equation}}

where
$$p\left(i_{t-1}({\underline l} )|i_1^n({\underline
l}),k_1^{t-1}({\underline l}),i_1^{t-2}({\underline l})\right)=
{\rm
{tr}}\left({\cal R}_{t-1,i_{t-1}({\underline l})}(i_1^{t-2}({\underline
l}),k_1^{t-1}({\underline l}))\left(\omega^{t-1}_{i_1^n({\underline
l})}(k_1^{t-1}({\underline l}),i_1^{t-2}({\underline
l})))\right)\right)$$
and
\small{\begin{equation}\begin{array}{ll}
p\left(k_t({\underline l} )|i_1^n({\underline
l}),k_1^{t-1}({\underline l}),i_1^{t-1}({\underline l})\right)&={\rm{tr}}
\left(\otimes_{j=1}^l\Phi^{(t)}(j)\right)\\
\;&\circ\left({\cal R}_{t-1,i_{t-1}({\underline l})}(i_1^{t-2}({\underline
l}),k_1^{t-1}({\underline l}))\left(\omega^{t-1}_{i_1^n({\underline
l})}(k_1^{t-1}({\underline l}),i_1^{t-2}({\underline
l}))\right)\right)\otimes_{j=1}^l F_{k_t(j)}^jF_{k_t(j)}^{j*}
\end{array}
\end{equation}}

 Once again,
we emphasise that we are conditioning here both on the initial
input state, indexed by $i_1^n({\underline l})$, and the sequence
of receiver measurements, indexed by the strings
$k_1^t({\underline l}),i_1^{t-1}({\underline l})$. Then we
calculate $$ p(i_1^n({\underline l})|k_1^{t}({\underline
l}),i_1^{t}({\underline l}))= \frac{ p\left(i_1^{t}({\underline
l}),k_1^{t}({\underline l})|i_1^{n}({\underline l})\right)
p(i_1^n({\underline l}) } {\sum_{i_1^{n}({\underline
l})}p\left(i_1^{t}({\underline l}),k_1^{t}(l)|i_1^{n}({\underline
l}))\right) p(i_1^n({\underline l})}. $$ The form of the
measurements ${\cal R}_j$ is then given by the POVM elements:

\begin{equation}\begin{array}{ll}R_{i_t(l)}(i_1^{t-1}({\underline
l}),k_1^t({\underline l}))&=\big(\sum_{r_t'\not= i_t(l)}
\Gamma_{r_t'}(k_1^t({\underline l}),i_1^{t-1}({\underline
l}))\big)^{-1/2}\notag\\ \,&\Gamma_{r_t}(k_1^t({\underline
l}),i_1^{t-1}({\underline l}))\big(\sum
\Gamma_{r_t'}(k_1^t({\underline l}),i_1^{t-1}({\underline
l}))\big)^{-1/2}\notag,
\end{array}\end{equation}

where the summation is over all $k_1^t({\underline
l}),i_1^{t-1}({\underline l})$ and $r_t\not =i_t({\underline l})$.
Furthermore

\begin{equation}\begin{array}{ll}\Gamma_{r_t}(k_1^t,i_1^{t-1}({\underline l}))
&=\Pi_{\rho^t,\delta}^l(k_1^t,i_1^{t-1}({\underline
l}))\Pi_{r_t}(k_1^t,i_1^{t-1}({\underline l}))\notag
\\
\, &\times\Pi_{\rho^t,\delta}^l(k_1^t,i_1^{t-1}({\underline
l}))\notag,
\end{array}\end{equation}
where $\rho^t$ is shorthand for
 $\rho^t(k_1^t({\underline l})),i_1^{t-1}({\underline l}))$, the ensemble
average of
$${\cal E}_t(k_1^t({\underline l})),i_1^{t-1}({\underline l}))$$
and the typical subspace projectors
$\Pi_{\rho^t,\delta}^l,\Pi_{r_t}(k_1^t({\underline l}
),i_1^{t-1}({\underline l}))$ (defined in the appendix) are
calculated with respect to ensemble ${\cal E}_t(k_1^t({\underline
l})),i_1^{t-1}({\underline l}))$ or the corresponding EHS quantum
state. The expected error probability on or before global round
$t$ (for error probability up to round $t$ denoted $P_t$) for this
protocol is evaluated inductively as follows, where we have
denoted the event that no error occurs on round $t$ by $E_t$:

\begin{equation}\begin{array}{ll}{\bf E}(P_t)&={\bf E}(P_{t-1})+{\bf
E}\left(P(\overline{E_t}|E_1,\ldots,E_{t-1})(1-P_{t-1})\right)\\
\, &={\bf
E}\left(P_{t-1}(1-P(\overline{E_t}|E_1,\ldots,E_{t-1}))\right)+{\bf
E}\left(P(\overline{E_t}|E_1,\ldots,E_{t-1})\right),
\end{array}\end{equation}
where we write $P(\overline{R_t}|R_1,\ldots,R_{t-1})$ for the
probability that an error occurs on round $t$ conditional on no
error before this point. The expectation is taken over all random
codes as constructed above. For $R\leq C$, from the argument given
in the appendix it follows that $${\bf E}(P_t)\leq {\bf
E}(P_{t-1})+\epsilon_t',$$ from which it is clear  that $${\bf
E}(P_t)\leq \sum_{k=1}^{t}\epsilon_k'.$$ Now the error probability
for the whole protocol is given by $${\bf E}(P_n)\leq \sum_{k=1}^n
k\epsilon_k'\leq (\sqrt{24}+6)\sum_{t=1} ^n
k\epsilon_k+\sum_{k=1}^n k2^{-t\;l\;K},$$ which $\to 0$ when we
let $l,n\to\infty$. Then by the standard argument, there a
(non-random) code with asymptotically $0$ error probability.

\section{Conclusion:} We have generalised the usual protocols for
classical communication using quantum channels, to include the
possibility of feedback. We proved a (regularised, asymptotic)
formula for the capacity using such protocols and showed that the
usual classical capacity formula \cite{holevo} follows from our
formula. The capacity formula is expressed in terms of the quantum
directed information, in an analogous sense to the classical
feedback capacity result \cite{tatikonda}. However, in a
significant departure from the classical case, it was noted that
the feedback capacity of a discrete memoryless channel may, at
least for some channels, exceed its unassisted capacity if one
allows the use of entangled input states.

\vskip 0.5 truecm {\bf Acknowledgements :} The author would like
to thank Yeo Ye and Yurii Suhov for useful and interesting
discussions. This publication is an output from project activity
funded by the Cambridge-MIT Institute limited ("CMI"). CMI is
funded in part by the United Kingdom government. The activity was
carried for CMI by Cambridge University and Massachusetts
Institute of Technology. CMI can accept no responsibility for any
information provided or views expressed.

\section{Technical Remarks}
In this section we will prove the estimates for the error
probabilities that were used in the preceding section. First we
require two lemmas:  firstly to accurately describe to what extent
 the measurements ${\cal R}_t$,
$1\leq t\leq n$, disturb the states they act on, and a second
lemma given estimates for quantities involving typical projectors.

\begin{lemma}
Let $\epsilon_t=2^{-tlc\delta^2}$ for some 
$c>0$ .
Then for $l$ sufficiently large, we have
$$||\omega_{i_1^n(1)}(k_1^t(1))\otimes\ldots\otimes\omega_{i_1^n(l)}(k_1^t(l)))-
\omega_{i_1^n({\underline l})}(i_1^{t-1}({\underline
l}),k_1^t({\underline l})
)||_1\leq\sum_{s=1}^{t}(\sqrt{24\epsilon_s}+6\epsilon_s), $$ 
for all $k$. The norm $|| A||_1$ is defined for all Hermitian $A$
in a finite dimensional space as the sum of the absolute values of
the eigenvalues of $A$. Here, the states
$\omega_{i_1^n(j)}(k_1^t(j))$, $1\leq j\leq n$ are defined
recursively as per the specifications of the $n$-block feedback
code ${\cal C}^n$.
\end{lemma}

\begin{proof}
This is done by induction. For $m=1$, denote the POVM elements of
${\cal R}_1$ as $R_{i_1({\underline l})}$, with the understanding
that we have "padded" the POVM elements by taking the tensor
product of the original elements with identity operators on those
copies of the channel space on which the measurement does not act.
 A similar convention is assumed
for ${\cal R}_2,\ldots,{\cal R}_n$. Then  we have $${\rm
{tr}}\left(\omega_{i_1^n(1)}(k_1(1))\otimes\ldots\otimes\omega_{i_1^n(l)}(k_1(l))
R_{i_1({\underline l})}\right)\geq 1-3\epsilon_1,$$ a consequence
of Lemma 6 of Hayashi and Nagaoka \cite{haya}. This implies
$$||\omega_{i_1^n(1)}(k_1(1))\otimes\ldots\otimes\omega_{i_1^n(l)}(k_1(l))-
\frac{\sqrt{R_{i_1({\underline
l})}}\omega_{i_1^n(1)}(k_1(1))\otimes\ldots\otimes\omega_{i_1^n(l)}(k_1(l))\sqrt{R_{i_1(l)}}}
{{\rm{tr}}\left(\omega_{i_1^n(1)}(k_1(1))\otimes\ldots\otimes\omega_{i_1^n(l)}(k_1(l))
R_{i_1({\underline l})}\right)}||$$ $$\leq
\sqrt{24\epsilon_1}+6\epsilon_1,$$

by an extension of the Winter tender measurement lemma
\cite{winter}.

Thus the required inequality holds in this case, noting that $$
\omega_{i_1^n({\underline l})}(i_1^{t}({\underline
l}),k_1^{t}({\underline l}) )= \frac{\sqrt{R_{i_1({\underline
l})}}\omega_{i_1^n(1)}(k_1(1))\otimes\ldots\otimes\omega_{i_1^n(l)}(k_1(l))\sqrt{R_{i_1(l)}}}
{{\rm
{tr}}\left(\omega_{i_1^n(1)}(k_1(1))\otimes\ldots\otimes\omega_{i_1^n(l)}(k_1(l))
R_{i_1({\underline l})}\right)}. $$

Assume that the statement holds for $m=t-1$. Then for $m=t$ we
have the following inequalities:

$$||\omega_{i_1^n(1)}(k_1^{t}(1)))\otimes\ldots\otimes\omega_{i_1^n(l)}(k_1^{t}(l)))-
\omega_{i_1^n({\underline l})}(i_1^{t}({\underline
l}),k_1^{t}({\underline l}) )||$$

$$\leq || {\cal G}_{k_t({\underline
l})}(\otimes_{j=1}^l\omega_{i_1^n(j)}(k_1^{t-1}(j)))
-{\cal G}_{k_t({\underline
l})}\left(\frac{\sqrt{R_{i_t({\underline
l})}} \otimes_{j=1}^l\omega_{i_1^n(j)}(k_1^{t-1}(j))\sqrt{R_{i_t({\underline l})}}}{{\rm
{tr}}\left(\otimes_{j=1}^l\omega_{i_1^n(j)}(k_1^{t-1}(j)) R_{i_t({\underline l})}
\right)}\right) ||$$

$$+||{\cal G}_{k_t({\underline
l})}\left(\frac{\sqrt{R_{i_t({\underline
l})}}\otimes_{j=1}^l\omega_{i_1^n(j)}(k_1^{t-1}(j))\sqrt{R_{i_t({\underline l})}}}{{\rm
{tr}}\left(\otimes_{j=1}^l\omega_{i_1^n(j)}(k_1^{t-1}(j))R_{i_t({\underline l})}
\right)}\right)-{\cal G}_{k_t({\underline
l})}\left(\omega_{i_1^n({\underline l})}(i_1^{t-1}({\underline
l}),k_1^{t-1}({\underline l}) ))\right)||$$
$$\leq\sum_{n=1}^{t}(\sqrt{24\epsilon_n}+6\epsilon_n),$$ using the
triangle inequality for the trace norm, the H\"older inequality
${\rm{tr}}(|A\;B|)\leq ||A||_1\;||B||_1$, the fact that the trace
distance is non-decreasing under completely positive operations
and the inductive hypothesis. Here the completely positive map $G$
is given by $${\cal G}_{k_t({\underline
l})}=\otimes_{j=1}^l {\cal N}^{k_{t-1}(j)}\otimes_{j=1}^l
{\cal M}_{t,k_t(j)}^j\circ
\left(\otimes_{j=1}^l\Phi^{(t)}(j)\right)\circ\otimes_{j=1}^l {\cal N}_{t,k_{t-1}(j)}$$
\end{proof}

What follows is an summary of some important definitions and
results concerning typical projectors (see \cite{winthesis} for a
more comprehensive discussion): Consider a general
classical-quantum system $UXQ$ of the form
$$\rho_{UXQ}=\sum_{u,x}p_{u,x}|(u,x\rangle\langle (u,x)|\otimes
\rho_{u,x},$$ where $u$ is defined on set $U$ and $x$ is defined
on set $X$. The set of typical sequences is defined by $${\cal
T}_{p,\delta}^n=\{x_1^n\; : \;\forall x\;|N(x|x_1^n)-np(x)|\leq
n\delta\},$$ where $N(x|x_1^n)$ counts the number of occurrences
of $x$ in the word $x_1^n=(x_1,\ldots,x_n)$.

For a density matrix $\rho=\sum_k \lambda_k |k\rangle\langle k|$,
define the probability distribution $P(K=k)=\lambda_k$ and for
$\delta>0$ the typical projector
$$\Pi_{\rho,\delta}=\sum_{k_1^n\in{\cal
T}_{K,\delta}^n}|k_1^n\rangle\langle k_1^n|.$$ Here we use the
shorthand $|k_1^n\rangle\langle k_1^n|=|k_1\rangle\langle
k_1|\otimes\ldots\otimes | k_n\rangle\langle k_n|$.

We also define the conditionally typical projector as:
$$\Pi^n_{\rho_u,\delta}(u_1^n)=\otimes_u
\Pi_{\rho_u,\delta}^{I_u},$$ where $I_u=\{i\; :\; u_i=u\}$ and
$\Pi_{\rho_u,\delta}^{I_u}$ is the typical projector of $\rho_u$
in those positions of the $n$-factor tensor product (representing
$u_1^n$) indicated by $I_u$.

This notation is slightly abused in the main part of this text,
where the conditional typical projector is written, for example,
as $$\Pi_{r_t}(i_1^{t-1}({\underline l}),k_1^t({\underline l})),$$
where $r_t=i_t'({\underline l})$ for some length $l$ string
$i_t'({\underline l})$. This refers to a projector (with respect
to a $l$ factor tensor product) of the form
$$\otimes_{j=1}^{l}\Pi_{\rho_{i_1^t(j),k_1^t(j),\delta}},$$ with
$\Pi_{\rho_{i_1^t(j),k_1^t(j),\delta}}$ the usual typical
projector.

With the above definitions, we have the following lemma: \vskip
0.5 truecm
\begin{lemma}
For typical projector $\Pi_{\rho,\delta}$, the
following relations hold: $${\rm {tr}}(\rho^{\otimes
n}\Pi_{\rho,\delta})\leq 2^{-n(S(\rho)+c\delta)}$$ and
$$\Pi_{\rho,\delta}\rho^{\otimes n}\Pi_{\rho,\delta}\leq
2^{-n(S(\rho)+c\delta)} \Pi_{\rho,\delta}$$ 
\end{lemma}
\begin{proof}
\end{proof}

We are now able to estimate the error probabilities, which we
evaluate as:
 $$P(\overline{E}_k|E_1,\ldots,E_{k-1})=1-{\rm
{tr}}(\omega_{i_1^n(1)}(k_1^t(1)))\otimes\ldots\otimes\omega_{i_1^n(l)}(k_1^t(l)))R_{i_t({\underline
l})})+\delta_{k-1},$$ where $\delta_{t-1}=\sum_{m=1}^{t-1}
(6\epsilon_{m}+\sqrt{24\epsilon_m})$. This follows from the above
Lemma. Then we have
 {\small\begin{equation}
\begin{array}{l}
1-{\rm
{tr}}(\otimes_{j=1}^l\omega_{i_1^n(j)}(k_1^{t}(j))R_{i_t({\underline
l})})\leq
2{\rm{tr}}\left(\otimes_{j=1}^l\omega_{i_1^n(j)}(k_1^{t}(j))\left(I-\Gamma_{i_t(l)}(k_1^{t}({\underline l}),i_1^{t-1}({\underline l}))\right)\right)\\
+4\sum_{r_t\not=i_t(l)}{\rm
{tr}}\left(\otimes_{j=1}^l\omega_{i_1^n(j)}(k_1^{t}(j))\Gamma_{r_t}(k_1^{t}({\underline
l}),i_1^{t-1}({\underline l}))\right).
\end{array}
\end{equation}}
by Lemma 6 in Hayashi and Nagaoka \cite{haya}. Furthermore we have
$${\rm
{tr}}\left(\omega_{i_1^n(1)}(k_1^t(1)))\otimes\ldots\otimes\omega_{i_1^n({\underline
l})}(k_1^t({\underline l})))
\Gamma_{i_t(l)}(k_1^t(l),i_1^{t-1}(l))\right)\geq 1-3\epsilon_t$$
by Lemma 6 in Hayashi and Nagaoka \cite{haya}. In addition we have

{\small
\begin{equation}
\begin{array}{l}
\sum_{k_1^t({\underline l})i_1^n({\underline l})}p_{i_1^n({\underline l})}\Pi_{j=1}^l
p(k_1^t(j)|i_1^n(j))
{\rm{tr}}\left(\otimes_{j=1}^l\omega_{i_1^n(j)}(k_1^t(j)))\Gamma_{r_t}(i_1^{t-1}(l),k_1^t(l))\right)\\
={\rm{tr}}\big((\sum_{i_1^n}p_{i_1^n}p(k_1^t|i_1^n)\omega_{i_1^n(l)}(k_1^t))^{\otimes
l}\Gamma_{r_t}(i_1^{t-1},k_1^t)\big) .
\end{array}
\end{equation}}

This in turn is equal to

$$ {\rm
{tr}}\bigg((\sum_{i_1^t,k_1^t}p_{i_1^t,k_1^t}\omega_{i_1^t,k_1^t})^{\otimes
l}\Pi_{\omega^t,\delta}^l(i_1^{t-1},k_1^t)
\Pi_{r_t,\delta}^l(i_1^{t-1},k_1^t)\Pi_{\omega^t,\delta}^l(i_1^{t-1},k_1^t)
\bigg) $$

$$ \leq 2^{-l H({\cal Z}_t|{\cal Z}_1^{t-1})+l c \delta}{\rm
{tr}}\left(\Pi_{r_t,\delta}^l(i_1^{t-1},k_1^t)\Pi_{\omega^t,\delta}^l(i_1^{t-1},k_1^t)\right)
,$$

for some constant $c>0$. This follows from the inequality
$$\Pi_{\omega^t,\delta}^l(i_1^{t-1},k_1^t)(\sum_{i_t,k_t}p_{i_1^t,k_1^t}\omega_{i_1^t,k_1^t})^{\otimes
l}\Pi_{\omega^t,\delta}^l(i_1^{t-1},k_1^t)\leq 2^{-l H( {\cal
Z}_t|{\cal Z}_1^{t-1})+l c
\delta}\Pi_{\omega^t,\delta}^l(i_1^{t-1},k_1^t)$$ Then
 we
have

$$ {\rm
{tr}}\left(\Pi_{r_t,\delta}^l(i_1^{t-1},k_1^t)\Pi_{\omega^t,\delta}^l(i_1^{t-1},k_1^t)\right)
\leq 2^{l H({\cal Z}_t|{\cal Z}_1^{t-1} A_1^t)-l c \delta}. $$

\vskip 0.5 truecm

Putting these estimates together we get $${\bf E}\left(1-{\rm
{tr}}(\omega_{i_1^{t}(1),k_{1}^{t-1}(1)}\otimes\ldots\otimes
\rho_{i_1^{t}(l),k_{1}^{t-1}(l)}R_{i_t(l)})\right)\leq 6\epsilon_t+4.2^{nl(\frac{1}{n}R(t)-\frac{1}{n}I(A_1^t:
{\cal Z}_t|{\cal Z}_1^{t-1}))} .$$ In exactly the same way we get
$$ {\bf E}\left(1-{\rm {tr}}(\omega_{i_1(1)}\otimes\ldots\otimes
\rho_{i_1(l)}R_{i_1(l)})\right)\leq 6\epsilon_1+4.2^{nl(\frac{1}{n}R(1)-\frac{1}{n}I(A_1:
{\cal Z}_1))},$$ where, for $1\leq t\leq n$, $R(t)=\frac{\log
N_t}{l}$, for $N_t$ defined as the number of strings
$i_t({\underline l})$ that label the measurement ${\cal R}_t$. We
necessarily have $R(1)+\ldots+R(n)=R$. We have assumed in this
paper that we consider only protocols for which the limit
$C=\lim_{n\to\infty}\frac{1}{n}I(A_1^n\rightarrow{\cal Z}_1^{n})$
exists. For any positive numbers $R(1),\ldots,R(n)$ as above
satisfying $R(t)\leq I(A_1^t: {\cal Z}_t|{\cal Z}_1^{t-1})$, we
clearly have $R\leq C$ and there exists $K>0$ such that for $n,l$
sufficiently large, we have

\begin{equation}\begin{array}{ll}
P(\overline{E}_t|E_1,\ldots,E_{t-1})&\leq
\sum_{m=1}^{t-1}(6\epsilon_m+\sqrt{24\epsilon_m})+6\epsilon_t+2^{-nlK}\\
\;&\leq
(\sum_{m=1}^t\epsilon_m)(6+\sqrt{24})+2^{-nlK}=\epsilon_t'.
\end{array}\end{equation}

\end{document}